\documentclass[aps,preprint,prl,double-spaced,showpacs,groupedaddress]{revtex4}  
\usepackage{graphicx}  
\usepackage{dcolumn}   
\usepackage{bm}        
\usepackage{amssymb}   
\usepackage{epsfig}

\begin{document}

\title{Efficient Monte Carlo Calculations of the One-Body Density}

\author{Roland Assaraf$^{a)}$, Michel Caffarel$^{b)}$, and Anthony Scemama$^{c)}$}

\affiliation{
$^{a)}$ Laboratoire de Chimie Th\'eorique, CNRS-UMR 7616,
Universit\'e Pierre et Marie Curie Paris VI,
Case 137, 4, place Jussieu 75252 PARIS Cedex 05, France\\
$^{b)}$ Laboratoire de Chimie et Physique Quantiques, CNRS-UMR 5626,
IRSAMC Universit\'e Paul Sabatier, 118 route de Narbonne
31062 Toulouse Cedex, France\\
$^{c)}$ Laboratoire Cermics, Ecole Nationale des Ponts et Chauss\'ees,
6 et 8 avenue Blaise Pascal, Cité Descartes-Champs sur Marne, 
77455 Marne la Vallée Cedex 2, France
}
\date{\today}

\begin{abstract}
An alternative Monte Carlo estimator for the one-body density $\rho({\bf r})$
is presented. This estimator has a simple form and can be readily used in any type of 
Monte Carlo simulation. Comparisons with the usual regularization of the delta-function  
on a grid show that the statistical errors are greatly reduced.
Furthermore, our expression allows accurate calculations of 
the density at any point in space, even in the regions never visited during 
the Monte Carlo simulation. The method is illustrated with the computation 
of accurate Variational Monte Carlo electronic densities for the Helium atom (1D curve) and for 
the water dimer (3D grid containing up to 51x51x51=132651 points).

\end{abstract}

\maketitle

The Monte Carlo approach is probably one of the most widely employed numerical approaches
in the scientific and engineering community. In computational physics, it 
has been extensively used in the last fifty years for studying a great variety 
of many-body systems under many different conditions. 
To date, the most popular application of the method is probably the 
calculation of classical thermodynamical properties.\cite{binder}
However, the Monte Carlo approach is also employed for evaluating quantum properties 
by using the Path-Integral formulation of quantum averages as classical ones (Quantum 
Monte Carlo or Path Integral Monte Carlo approaches\cite{cep}). 
In the recent years, these later approaches have emerged as an unique and powerful tool 
for studying quantitatively the interplay between quantum and thermal effects 
in many-body systems ({\it e.g.}, to understand the very rich physics 
of strongly correlated materials).

At the heart of all these applications lies the calculation of a number of high-dimensional
integrals (or sums, for lattice problems) written under the general form
\begin{equation}
I(F) = \int d{\bf r}_1 \cdots d{\bf r}_N \Pi({\bf r}_1, \cdots,{\bf r}_N) 
F({\bf r}_1,\cdots, {\bf r}_N) 
\label{eq1}
\end{equation}
where $\Pi$ is some arbitrary $N$-body probability distribution ($\Pi$ positive and 
normalized) and $F$ some arbitrary real-valued function. The integration is performed
over all accessible configurations for the $N$-particle system. The general idea of Monte Carlo 
approaches is to evaluate the integral by sampling the configuration space according to 
the probability distribution, $\Pi$, and by averaging $F$ over 
the various configurations generated by the sampling procedure,
$I(F) = \langle F \rangle_{\Pi}.$
Here and in what follows, the symbol $\langle \cdots \rangle_{\Pi}$ 
indicates the statistical average over the density $\Pi$. Various Monte Carlo 
algorithms (sampling procedures) can be found in the literature, the most celebrated 
one being, of course, the Metropolis algorithm.\cite{metro}
The efficiency of a Monte Carlo approach is directly related to the magnitude of 
the fluctuations of the integrand in the regions where the probability distribution,
$\Pi$, is large. More precisely, for a given number of Monte Carlo steps, the statistical 
error $\delta F$ is proportional to the square root of the variance of 
the integrand $F$ defined as $\sigma^2(F) \equiv I(F^2)-I(F)^2$.
Accordingly, a very attractive way of enhancing the convergence of a Monte Carlo 
simulation consists in introducing alternative ``improved'' estimators defined as 
new integrands $\tilde{F}$ having the same average as $F$ but a lower variance 
\begin{equation}
\langle \tilde{F} \rangle_{\Pi} = \langle F \rangle_{\Pi} 
\;\;\;{\rm and} \;\;\; \sigma^2(\tilde{F}) < \sigma^2(F).
\end{equation}
In previous works\cite{zv1,zv2} it has been shown how improved
estimators can be designed for any type of integrand $F$ and
Monte Carlo algorithm, and some applications to the computation of forces have been 
presented\cite{zv2}. 

In this Letter we present an efficient improved Monte Carlo estimator for calculating
the one-body (or one-particle) density, $\rho({\bf r})$
\begin{equation}
\rho({\bf r})=
\langle
\sum_{i=1}^{N} \delta({\bf r}_i-{\bf r})
\rangle_{\Pi}
\label{defrho}
\end{equation}
and, more generally, any one-body average of the form $\int d{\bf r} \rho({\bf r}) F({\bf r})$. 
As we shall see, our estimator allows
very important reductions in variance. In the example of the charge density of 
the water dimer presented below, a reduction of up to 
two orders of magnitude in CPU time is possible for some regions of space.
In addition, and in sharp constrast with the usual estimator based on the regularization of the 
delta-function on a grid, our expression leads to accurate estimates of
the density at any point in space, even in the regions never visited during
the Monte Carlo simulation ({\it e.g.}, in the large-distance regime). 
This property is particularly interesting when a global knowledge of the density map 
is searched for. For the water dimer case, we were able to accurately compute the charge density 
for 51x51x51=132651 grid points. Note that such a calculation
is vastly more difficult to perform when using the standard approach.

Let us recall that accurate one-particle properties are 
of central interest for the understanding of the physics of many complex many-body systems. 
Such systems include all those which are not translationally invariant 
(typically, all finite systems: atoms, molecules, clusters, nuclei, etc.) and all those 
whose translational symmetry has been explicitly broken, e.g., by the application of 
an inhomogeneous external field. In addition to this, 
many physical modelizations and/or effective theories rely explicitly on the knowledge of 
the one-body density. Many examples could be cited but let us mention, for example,
the various modelings of the electrostatic field of molecules from 
the one-electron density\cite{gadre}, the studies of the structure and reactivity of molecular 
systems based on the topological analysis of the electron density and/or its 
Laplacian,\cite{bader}, and, also, the very important case of
Density Functional Theories (DFT) which could greatly benefit 
from the possibility of computing  accurate 3D charge/spin density maps 
for large molecular systems (e.g., via accurate fits of the exchange-correlation 
Kohn-Sham potential, see \cite{dft}).

Finally, let us note that the use of alternative forms for evaluating the
density is not new. For example, in the works of Hiller {\it et al.}\cite{hill},
Sucher and Drachman\cite{sucher}, Hariman\cite{hari}, Rassolov and Chipman\cite{rasso}
new classes of global operators built for computing the density have been introduced.
However, in these works, the general idea is to design operators whose expectation values
give an accurate estimate of the unknown exact one-body density 
and not the exact one-body density associated with a known
$N$-body density. Actually, our strategy is more closely related to what
has been presented by Vrbik {\it et al.}\cite{vrbik}, Langfelder {\it et al.}\cite{vrbik2}, and
Alexander and Coldwell\cite{alex}. In these works,
alternative Monte Carlo estimators with lower
variances are also introduced. However, the emphasis is only put on the case
of evaluating the charge and/or spin density at the nuclei.
Here, such ideas are extended to any point in space and a general formula
allowing to control all possible sources of statistical fluctuations in all possible 
regimes is presented.

{\it General improved density estimator}. 
Due to the presence of Dirac functions in the Monte Carlo estimator of the density, 
Eq.(\ref{defrho}), some sort of regularization has to be introduced. 
It is usually done by partionning the physically relevant part of the one-particle space 
(usually, the 3D ordinary space) into small domains of finite volume and by evaluating 
the corresponding locally-averaged densities. 
In practice, such a procedure is particularly simple to implement by counting
the number of particles present in each elementary domain at each step of the simulation.
However, the statistical fluctuations can be rather large.
This is particularly true for the low-density regions which are rarely 
visited by the particles. Even worse, there is no way of evaluating the 
density in regions which are {\it never} visited during the finite Monte Carlo simulation. 
One way to escape from these difficulties is to introduce some global estimators 
defined in the whole space. To do that, we regularize the $\delta$ Dirac-function 
by using the following equality
\begin{equation}
  \delta ({\bf r}_i - {\bf r}) =
 \frac{-f({\bf r}_i;{\bf r})}{4 \pi} {\bf \nabla}_i^2 \frac{1} {|{\bf r}_i - {\bf r}|},
\end{equation}
where $f({\bf r_i};{\bf r})$ is a smooth function of ${\bf r_i}$ 
(here, ${\bf r}$ plays the role 
of an external parameter) verifying $f({\bf r}_i={\bf r};{\bf r})=1$. Note that this formula 
is just a slightly generalized form of the well-known equality corresponding to $f=1$.
Now, injecting this expression into Eq.(\ref{defrho}) and integrating by parts, the density 
can be rewritten as
\begin{equation}
\rho({\bf r})=
-\frac{1}{4 \pi} \sum_{i=1}^{N} \langle 
\frac{1} {|{\bf r}_i - {\bf r}|}   
\frac{ {\bf \nabla}_i^2 (f \Pi)}{\Pi}
\rangle_{\Pi}.
\end{equation}
Next, we introduce some additional function $g({\bf r})$ independent on the particle 
coordinates and write $\rho({\bf r})$ under the form
\begin{equation}
\rho({\bf r})=
-\frac{1}{4 \pi} \sum_{i=1}^{N} \langle
[\frac{1} {|{\bf r}_i - {\bf r}|} - g]
\frac{ {\bf \nabla}_i^2 (f \Pi)}{\Pi}
\rangle_{\Pi},
\label{basic}
\end{equation}
the last step being allowed since $\langle \frac{{\bf \nabla}_i^2 (f \Pi)}{\Pi} \rangle_{\Pi}=0$. 

Expression (\ref{basic}) is our general form for the improved estimator of the one-body 
density. 
The two functions $f$ and $g$ play the role of auxiliary quantities. They are introduced to
decrease the variance of the density estimator. 
As with any optimization problem, there is no universal strategy for choosing $f$ and $g$.
However, the guiding principle is to identify the leading sources of fluctuations and, then, 
to adjust the auxiliary functions to remove most of them.

{\it Improved electronic density estimator for molecules}.

In what follows, we consider the one-electron
density of molecular systems issued from the $N$-body 
quantum probability distribution written as
\begin{equation}
\Pi = \frac{ \psi_T^2({\bf r}_1, \cdots,{\bf r}_N) }
           {\int d{\bf r}_1 \cdots d{\bf r}_N \psi_T^2({\bf r}_1, \cdots,{\bf r}_N)},
\label{}
\end{equation}
where $\psi_T$ is some electronic trial wavefunction.

1. {\it Short electron-nucleus distance regime}. 

For a system of electrons in coulombic interaction with a set of fixed nuclei 
the exact wave function is known to obey the following electron-nuclear cusp
condition
\begin{equation}
\psi \sim_{{\bf r}_i \rightarrow {\bf R}_A}  1 - Z_A |{\bf r}_i -{\bf R}_A |,
\end{equation}
where ${\bf R}_A$ denotes the position of a given nucleus $A$ of charge $Z_A$. 
Most of the accurate trial wavefunctions employed in the literature fulfill this 
important condition. Now, as a consequence of the cusp condition we have
\begin{equation}
\frac{ {\bf \nabla}_i^2 \Pi}{\Pi} 
\sim_{{\bf r}_i \rightarrow {\bf R}_A}  -\frac{4 Z_A}{|{\bf r}_i -{\bf R}_A |}.
\end{equation}
In the neigbhorhood of nucleus $A$, this term is an important source 
of fluctuations. This is easily seen by noting that, in the regime ${\bf r} 
\sim {\bf R}_A$, the estimator of 
$\rho({\bf r})$, Eq.(\ref{basic}), behaves as 
$\frac{1}{|{\bf r}_i -{\bf R}_A |^2}$, a quantity which 
has an infinite variance [$\int (\frac{1}{r^2})^2 r^2 dr =+\infty $].
To remove this source of wild fluctuations we adjust the function $f$ so that 
$\frac{ {\bf \nabla}_i^2 f}{f}$ exactly vanishes the divergence of $\frac{{\bf \nabla}_i^2 \Pi}{\Pi}$. 
A simple suitable form for $f$ verifying such a condition plus the 
constraint, $f({\bf r}_i={\bf r};{\bf r})=1$, is given by
\begin{equation}
f({\bf r_i};{\bf r}) = 1 +2 Z_A (|{\bf r}_i -{\bf R}_A|-|{\bf r}-{\bf R}_A|).
\label{eq11}
\end{equation}

2. {\it Large-distance regime}.

In the large-distance regime,$|{\bf r}| \rightarrow +\infty$, the exact
one-electron density is known to decay exponentially. In contrast, our basic estimator 
decays algebraically as a function of ${\bf r}$. Here, we propose to force the latter estimator
to decay also exponentially. 
In practice, a simple choice for the function $f$ is 
\begin{equation}
f({\bf r_i};{\bf r}) = (1 +\lambda |{\bf r}_i -{\bf r}|) \exp{[-\lambda |{\bf r}_i -{\bf r}|]},
\label{eq12}
\end{equation}
where $\lambda$ is some real parameter.
Note that the coefficients of the linear and exponential terms have been taken identical ($= 
\lambda$) to avoid the divergence of the laplacian of $f$ at ${\bf r}_i={\bf r}$. The  
parameter $\lambda$ can be adjusted eiher by minimizing the fluctuations of the average density 
or fixed at some value close to the theoretical value of $\lambda_{ex}=2\sqrt{-2I}$ 
($I$ first ionization potential, the exact density is known to decay as 
$\rho \sim \exp{-\lambda_{ex} r}$). 

Another useful remark is that, in the regime $|{\bf r}| \rightarrow +\infty$, the electrons 
of the molecule are, in first approximation, at the same distance from the point where 
$\rho$ is evaluated. As a consequence, during the simulation
the quantity $1/|{\bf r}_i -{\bf r}|$ fluctuates very little around the approximate average
$1/|\langle {\bf r}_i\rangle -{\bf r}|$.
Therefore, to reduce the fluctuations it is valuable to remove a quantity close to the 
latter average from the former one.
Here, this idea is implemented by introducing the following function $g$
\begin{equation}
g({\bf r}) = \frac{1}{M} \sum_{A=1}^M  \frac{1}{|{\bf R}_A -{\bf r}|}.
\label{eq13}
\end{equation}

\begin{center}
\begin{figure}[htp]
\includegraphics[width=0.8\linewidth,angle=0]{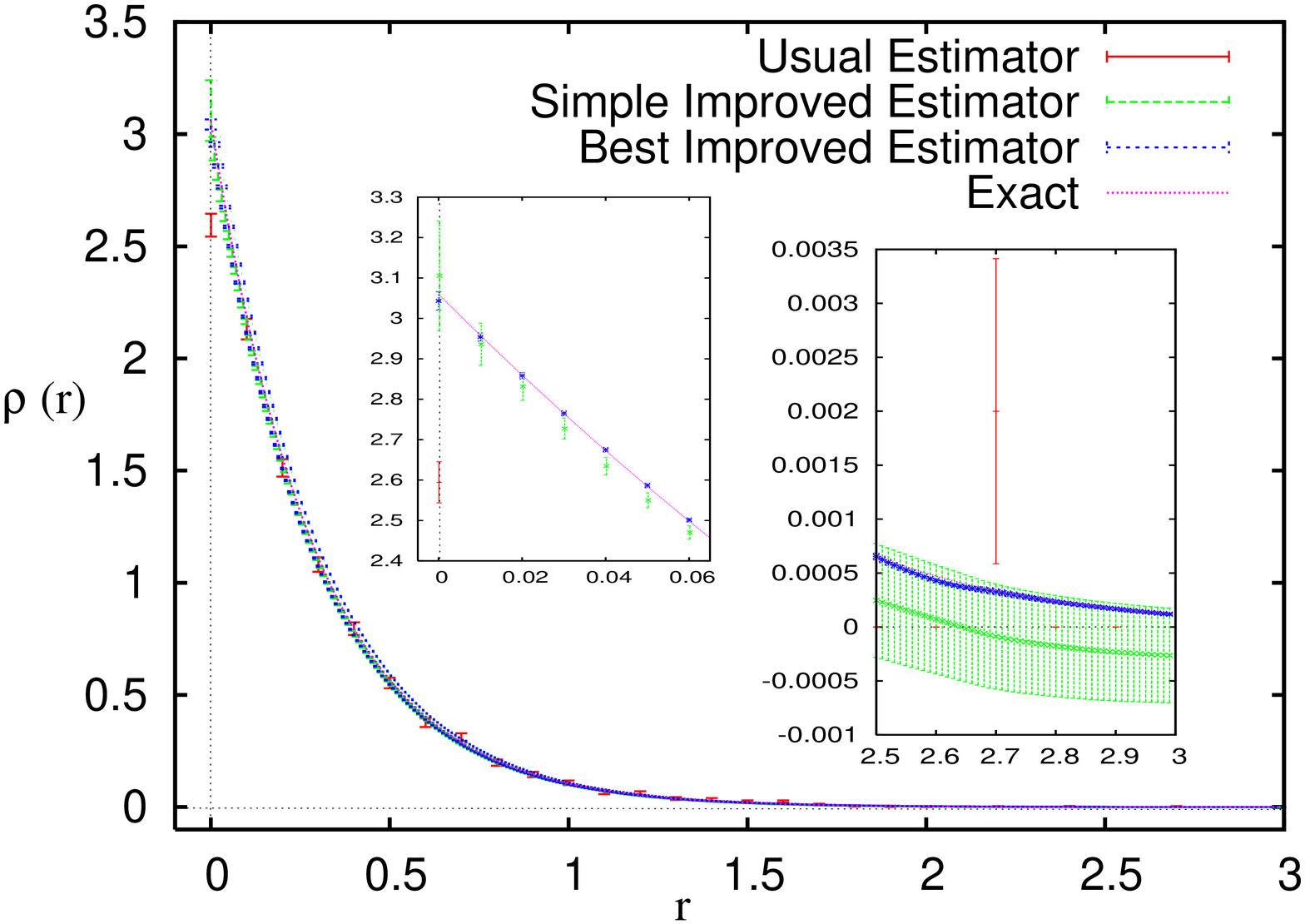}
\caption{ }
\label{fig1}
\end{figure}
\end{center}

In our first application we consider the He atom described by a simple trial wavefunction 
written under the form, $\psi_T=\phi({\bf r}_1)\phi({\bf r}_2)$ 
with $\phi= \exp{(-\gamma r)}$, and 
$\gamma=1.6875$ (Slater value). For this problem, the exact density is known and is given by 
$\rho(r)= 2\gamma^3/\pi \exp{(-2\gamma r)}$.
Figure \ref{fig1} shows the results obtained for $\rho(r)$ 
for a relatively short Monte Carlo run. 
The main curve displays the results obtained with i.) the usual estimator based on the delta 
representation, Eq.(\ref{defrho}), ii.) the simple improved estimator corresponding to 
Eq.(\ref{basic}) with $f=1$ and $g=0$, and our best improved estimator defined via 
Eqs.(\ref{basic},\ref{eq11},\ref{eq12},\ref{eq13}). In the latter case, the densities
corresponding to each of the two possible choices for $f$ [Eqs.(\ref{eq11}) or (\ref{eq12})] have 
been computed for each distance, the final value corresponding to the value having the smallest 
statistical error. 
The usual estimator, Eq.(\ref{defrho}), has been regularized by introducing small 
elementary cubes of length $a=0.2$. 
For all distances the statistical 
error associated with the usual estimator is very large with respect to improved estimators.
At intermediate distances, at least one order of magnitude in accuracy is lost.
For example, at $r=0.6$ the statistical error is about 10 times larger than for the simple 
estimator case and a factor of about 20 is found with respect to the best improved estimator. 
At large distances, a region rarely visited by the electrons, the standard estimator is so noisy 
that it is useless in practice. Now, regarding improved estimators it is clear that 
the auxiliary functions $f$ and $g$ have a great impact in reducing the errors.
At very small distances (first inset) a gain of about 5 in statistical error is 
obtained with the best improved estimator. At $r=0$, the gain is even larger since the 
simple estimator has an infinite variance.
At large distances where both improved estimators 
have a finite variance, it is seen that introducing some exponential decay into 
the estimator plus a proper shift ($g$-contribution) improves considerably the convergence. 
In the range (2.5,3.) the gain in error increases from 15 at $r=2.5$ to 40 at $r=3.$

\begin{center}
\begin{figure}
\includegraphics[width=0.8\linewidth,angle=0]{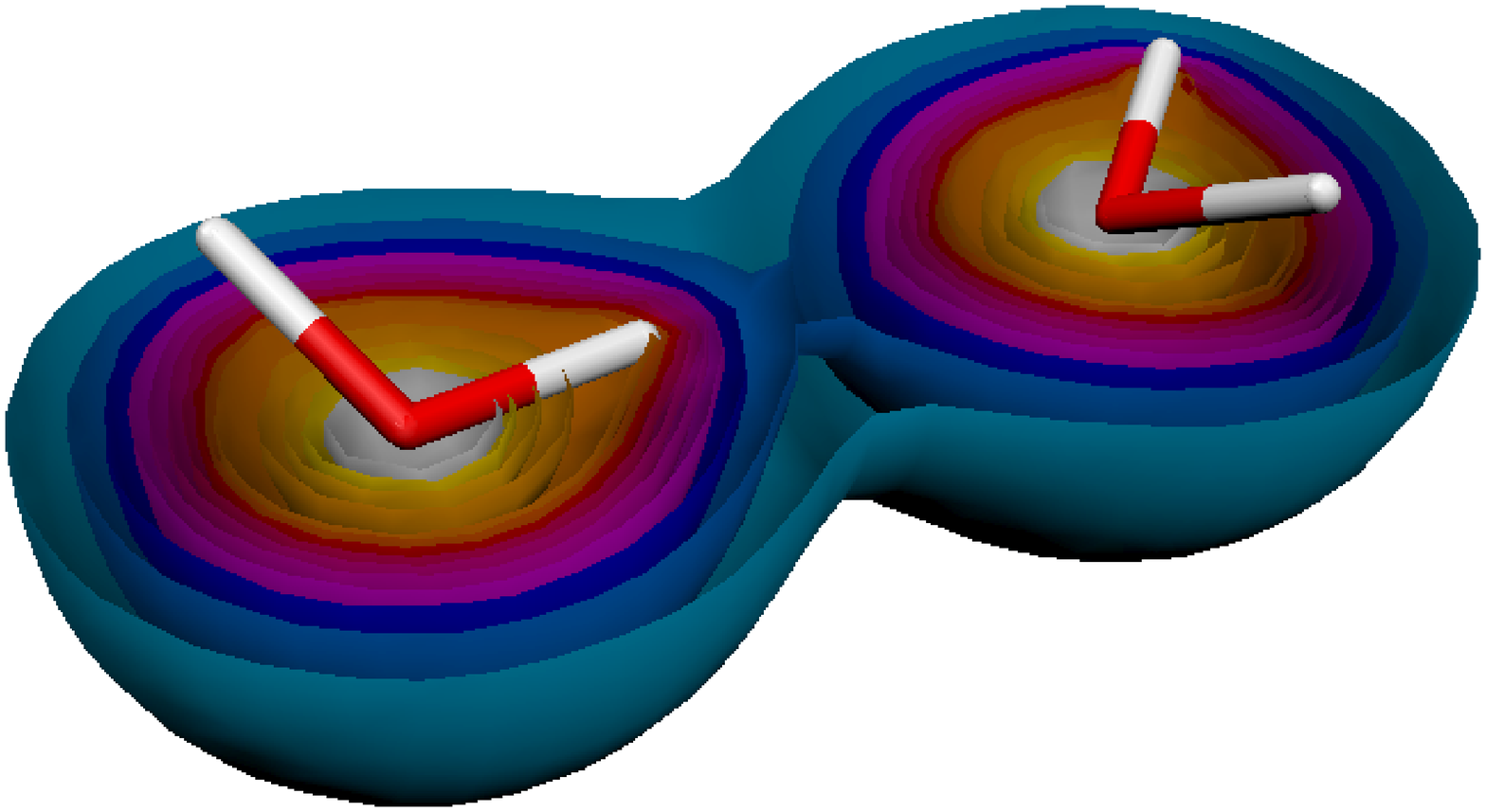}
\caption{}
\label{fig2}
\end{figure}
\end{center}

In our second application we consider the water dimer in a non-symmetric nuclear geometry 
(structure $\#$2 of Ref.\cite{tschumper}) described by an electronic wavefunction 
consisting of a Hartree-Fock part (cc-pVTZ basis set) plus a standard explicitly correlated 
Jastrow term. Figure \ref{fig2} shows the density plots obtained with our best
improved estimator for a number of points equal to 51x51x51 (=132651).
As seen on the figure the density obtained displays a very smooth aspect. 
A closer look shows that this regularity is present at a rather small scale. 
In Figure \ref{fig3} we present a more quantitative comparison of the data 
along the $O$-$O$ axis. The figure clearly shows that the best estimator 
outperforms the usual one. First, the curve corresponding to the 
new estimator (solid line connecting the points) 
is very smooth, although it has been obtained by simple linear extrapolation of the data. 
In sharp contrast, this is absolutely not true for the usual 
estimator curve whose overall behavior is particularly chaotic. Second, 
the statistical error has been greatly reduced using the new estimator. 
Depending on the distance, a gain in accuracy ranging roughly from 5 to 10 (i.e, up to 
two orders of magnitude in CPU time) has been obtained. An interesting point to mention 
is the presence of some very wild fluctuations in the neighborhood of r$\sim 1.5$ 
for the standard estimator. These fluctuations are due to 
the presence of a hydrogen atom close to the $O-O$ axis. We can verify that, in sharp contrast,
our new estimator, which has been built to correctly take into account the nuclear cusp, 
Eq.(\ref{eq11}), performs well in that region. Finally, remark that
in the large-$r$ regime (data not shown here) where the standard estimator is strictly zero 
(no sampling of this region), the improved estimator still continues to give accurate 
values of the very small density.

{\it Acknowledgments}
The authors would like to thank the ACI: ``Simulation Moléculaire'' for its support. Numerical
calculations have been performed using the computational ressources of
IDRIS (CNRS, Orsay) and CALMIP (Toulouse).




\newpage
\begin{center}
FIGURE CAPTIONS
\end{center}

\begin{itemize}
\item Fig.\ref{fig1} Density of Helium from various estimators, see text.

\item Fig.\ref{fig2} One-electron density of the $H_2O$ dimer with the
best improved estimator.

\item Fig.\ref{fig3} Cut of the one-electron density along the $O$-$O$ 
axis of the $H_2O$ dimer. Data for the
best and usual estimators. Solid lines are simple linear extrapolations of the data.

\end{itemize}

\newpage

\newpage

\begin{center}
\newpage
\begin{figure}
\includegraphics[width=0.8\linewidth,angle=0]{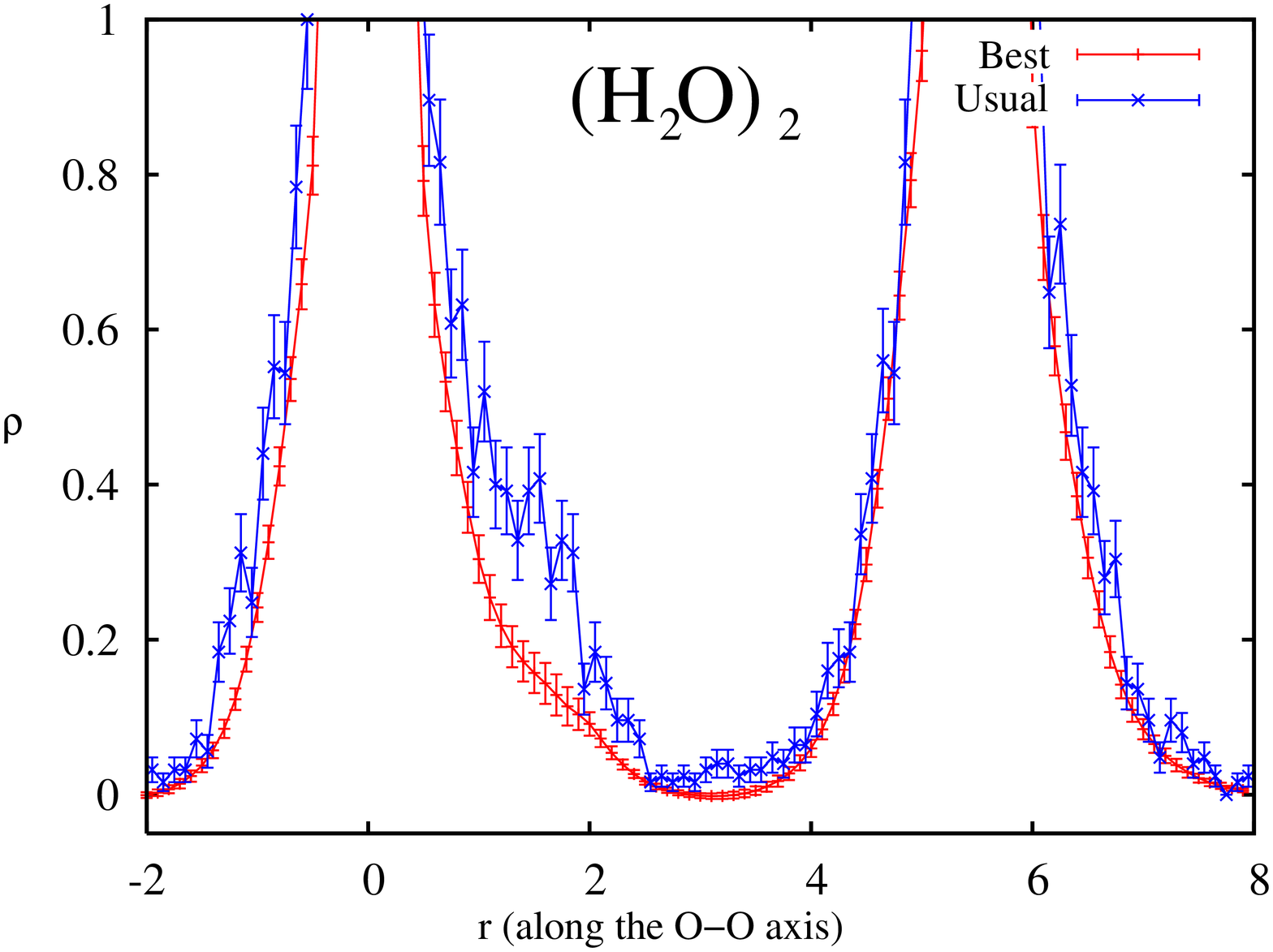}
\caption{}
\label{fig3}
\end{figure}
\end{center}

\end{document}